\documentclass[preprint2]{aastex}
\usepackage{ulem,color}

\slugcomment{to appear in the ApJ}

\shorttitle{A 3D model for the pPN CRL618}
\shortauthors{Vel\'azquez et al.}

\begin{document}


\title{An asymmetric jet launching model for the protoplanetary nebula CRL
618}

\author{Pablo F. Vel\'azquez}
\affil{Instituto de Ciencias Nucleares, Universidad Nacional
  Aut\'onoma de M\'exico, Apdo. Postal 70-543, CP: 04510,
  D.F., Mexico}
\email{pablo@nucleares.unam.mx}

\author{Angels Riera\altaffilmark{1}}

\affil{Departament de F\'\i sica i Enginyeria Nuclear, EUETIB, Universitat Polit\`ecnica de Catalunya, Comte d'Urgell 187, 08036 Barcelona, Spain }

\altaffiltext{1}{Departament d'Astronomia i Meteorologia, Universitat de Barcelona, Av. Diagonal 647, 08028 Barcelona, Spain}

\author{Alejandro C. Raga}
\affil{Instituto de Ciencias Nucleares, Universidad Nacional
  Aut\'onoma de M\'exico, Apdo. Postal 70-543, CP: 04510,
  D.F., Mexico}

\and 

\author{Juan C. Toledo-Roy}
\affil{Instituto de Ciencias Nucleares, Universidad Nacional
  Aut\'onoma de M\'exico, Apdo. Postal 70-543, CP: 04510,
  D.F., Mexico}

\begin{abstract}
  We propose an asymmetrical jet ejection mechanism in order to model
  the mirror symmetry observed in the lobe distribution of some
  protoplanetary nebulae (pPNe), such as the pPN CRL 618. 3D
  hydrodynamical simulations of a precessing jet launched from an
  orbiting source were carried out including an alternation in the
  ejections of the two outflow lobes, depending on which side of the
  precessing accretion disk is hit by the accretion column from a
  Roche lobe-filling binary companion. Both synthetic optical emission
  maps and position-velocity (PV) diagrams were obtained from the
  numerical results with the purpose of carrying out a direct
  comparison with observations. Depending on the observer's point of
  view, multipolar morphologies are obtained which exhibit a mirror
  symmetry at large distances from the central source. The obtained
  lobe sizes and their spatial distribution are in good agreement with
  the observed morphology of the pPN CRL 618. We also obtain that the
  kinematic ages of the fingers are similar to those obtained in the observations.
\end{abstract}

\keywords{ISM: jets and outflows---methods: numerical---planetary nebulae: individual (CRL
  618)}

\section{Introduction}

Explaining the multipolar morphology observed in proto planetary
nebulae (pPNe) represents a challenge. Several mechanisms have been
proposed in order to model the peculiar morphology exhibited by
these objects.
                                                                          
Some of these mechanism are based on the hypothesis that the central
source of the pPNe is actually a binary system. This idea was firstly
invoked by \citet{bond78} \citep[also][]{livio79,soker94}. The
existence of a binary system inside of the pPN can produce collimated
outflows when one of the components of the binary system, probably a
white dwarf, accretes material from its companion, an AGB or post-AGB
star \citep[e.g.][]{morris87,soker00}.  Subsequently, an accretion
disk is formed and a bipolar jet or collimated fast wind (CFW) is
ejected perpendicular to the plane of the disk
\citep{frankblackman04,frank07}.

The pioneering 3D hydrodynamical simulations of \citet{cliffe95} showed
that a precessing jet with a time-dependent ejection velocity can result in a
point-symmetric nebula.

Following this idea, high-resolution 3D hydrodynamical simulations of
a precessing jet with a time-dependent velocity were carried out in
order to model the bipolar morphology of the pPNe Hen 3-1475
\citep{velazquez04,riera04} and IC 4634 \citep{guerrero08}. Also, a
precessing and continuous jet was used to reproduce both the
morphology and emission of the PN K 3--35 \citep{velazquez07} and the
Red Rectangle \citep{vel11}.

Mirror and point-symmetric morphologies obtained from a precessing and
continuous jet emanating from a binary
system with circular orbits were studied by
\citet{raga09}. For orbital periods less than the precession
period, these authors found that mirror symmetry is observed
close to the central source, while point-symmetric morphologies
prevail at larger distances \citep{masciadri02,raga09}. The influence
of the orbital motion on the line emission of these objects was
analyzed by \citet{haro09}. They found that the apparent
mirror or point-symmetric morphology depends on the orientation
of the object with respect to the observer.

The pPN CRL~618 exhibits a four-lobe mirror symmetric morphology.
Numerical simulations have been carried out in the past with the intent of
reproducing both the morphology and kinematics of this pPNe, such as the study of
\citet{lee03} who proposed a scenario in which a CFW is interacting
with a spherical AGB wind \citep[see also][]{lee09}. Alternatively,
\citet{dennis08} proposed that the multiple jet appearance of CRL~618
could be due to clumps moving outwards at high velocities and slightly
different directions. \citet{balick13} have explored the nature of the
jets of CRL~618 by means of 3D simulations of either a
bullet or a continuous jet moving through the remnant AGB wind. These
authors favour the "bullet'' hypothesis based on the multipolar
morphology and the behaviour of the proper motion, which
  increases linearly with the distance from the central source.
   However, all these studies focus on modeling a single lobe.

By means of 3D hydrodynamical simulations, \citet{vel12} \citep[see
  also][]{vel13} have shown that the existence of a binary source
where one of the stellar components (the primary star) ejects
a spherical wind, while the orbiting companion (in elliptical orbit) emits a
bipolar, precessing jet with time-dependent
velocity, can generate multipolar geometries with
results similar to the observed features in pPNe.  These
authors found that the large-scale morphological characteristics of
these nebulae (lobe size, semi-aperture angle, number of observed
lobes) can be related to some of the parameters of the binary system.

\citet{soker13} compare the morphologies of the pPN CRL618 and
  the young stellar object (YSO) NGC 1333 IRAS 4A2, and propose that
  their morphologies can be explained by means of ``twin-jets'',
  e.g. two very narrow and nearby jets launched at the same time. They
  also consider that the origin of the jets is a binary system, with the
  components orbiting in very eccentric elliptical orbits. The ``twin-jets''
  are emitted near periastron passages.
 
Recently, \citet{riera14} carried out an observational and numerical
study of the pPN CRL~618. While the numerical model is based on the
work of \citet{vel13}, a time-decreasing trend was added to the jet
velocity in the model description in order to guarantee consistency
with the proper motion determined observationally by
\citet{riera14}. In spite of this, they obtain point-symmetric
morphologies in their synthetic maps.

In this work, we have added a new ingredient to our description:
  an asymmetric jet ejection, i.e., we either let both the jet and
  counter-jet be launched simultaneusly, or we launch just one of them,
  depending on whether the accretion disk is close to being edge-on from
  the perspective of the companion star when the system pass through
  its periastron. This mechanism could explain the lobe distribution
  observed in the pPN CRL~618. This work is organized as follows: the
assumptions of our model are given in section 2; the initial numerical
setup is given in section 3; in section 4 we list our results; and
finally in section 5 we summarize and discuss our results.

\section{Model assumptions}

Following the previous work of \citet{raga09,vel12,vel13} and \citet{riera14}, we
consider that a binary system is located at the centre of the pPN. One
of the components of this binary system accretes material from
the companion and launches a bipolar jet. The jet axis
changes in time, forming a precession cone with a
semi-aperture angle $\alpha$. The precession is retrograde with
respect to the orbital motion \citep[following the work
  of][]{terquem99}.

We assume that we have a jet with a periodic ejection velocity, with a
period equal to the orbital period \citep{vel12}. Additionally, we
impose a linear trend of decreasing ejection velocity with time in
order to reproduce the observed behavior of the proper motion velocity
\citep{riera14}.

In contrast to the scenario explored by \citet{riera14}, in this work we
consider an ``asymmetrical jet ejection mechanism'' which is inspired
by the results of \citet{montgomery12} and \citet{fendt13}. The first
of these papers presents a numerical study (by means of SPH
simulations) of precessing acretion disks. \citet{montgomery12} shows
that because of the precession the accreting material falls only on
one side of the accretion disk, producing asymmetries, or ``humps'',
in the density distribution. \citet{fendt13} studied how various disk
perturbations can produce an asymmetrical jet ejection which can last
for a few orbital periods.  Several asymmetries have been reported in
jets seen in YSOs. \citet{hirth94} performed a survey of T Tauri stars
and found that about 50\% of bipolar outflows observed in the optical
exhibit differences between the velocities of the blueshifted and
redshifted lobes that amount to factors of 1.5--2.5.

Recently, \citet{matsakos12} carried out both
magnetohydrodynamic and resistive magnetohydrodynamics numerical
simulations in order to explore intrinsic and extrinsic mechanisms
which could explain the differences in the velocities observed in the
blue and red components of YSO jets. The intrinsic mechanism is
related to the configuration of the magnetic field of the central
object, while the extrinsic one refers to the propagation of the jet
into an inhomogeneous circumstellar medium (CSM). As an example of
asymmetries due to the CSM we can mention the work of \citet{podio11},
which is based on observations of the DG Tauri B jets.

\begin{figure}
\epsscale{1.}
\plotone{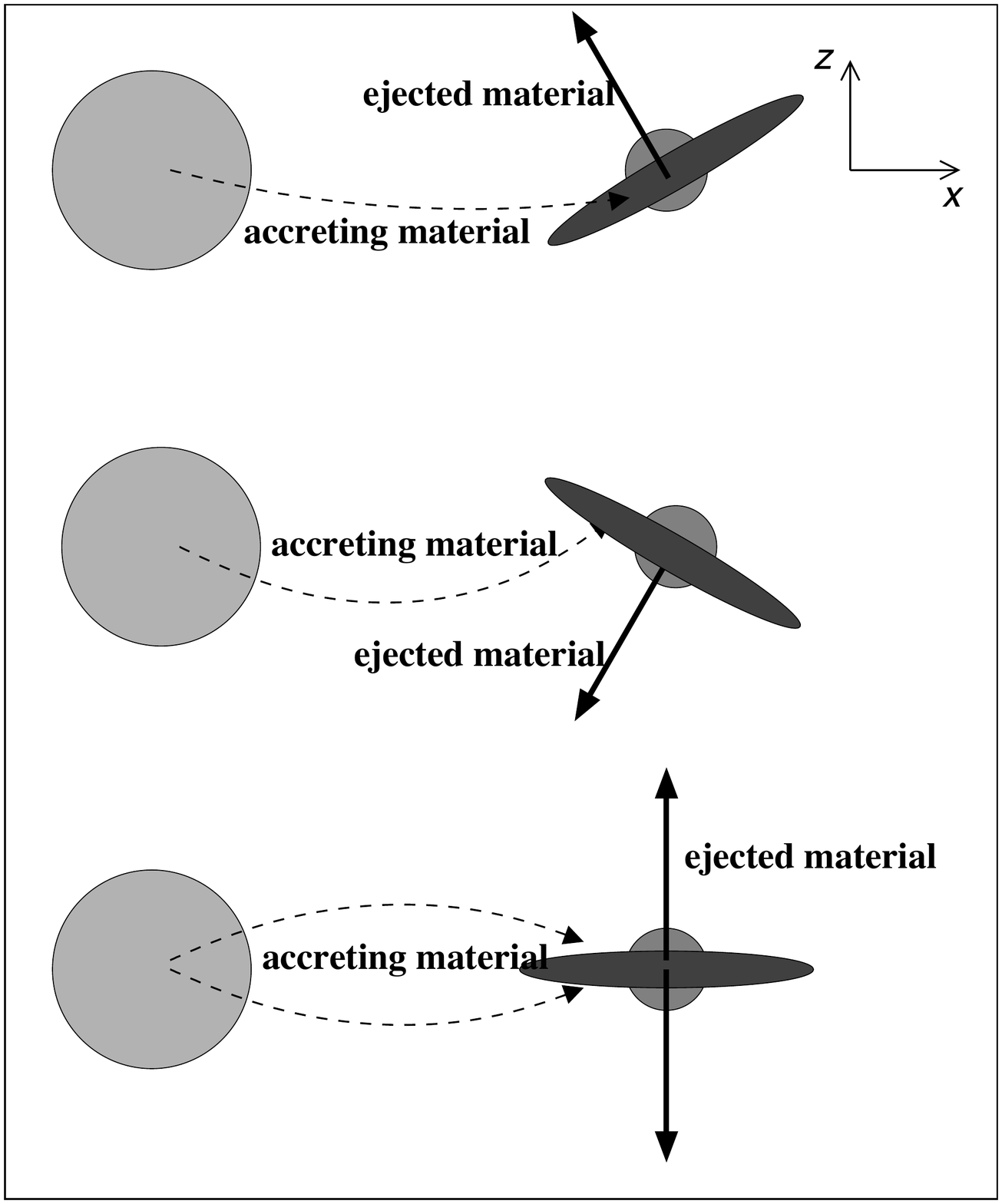}
\caption{Schematic diagram of the ejection condition. The orbit lies on the $xy$--plane (the $y$-axis is perpendicular to the plane of the figure) and both foci of the
  elliptical orbit are on the $x$--axis.
\label{fcroquis}}
\end{figure}

The asymmetrical jet ejection mechanism proposed in this work is that
the jet (or counter-jet) material is launched at periastron only
when that side of the accretion disk is subject to an infall of
accreted material, i.e., the jet (or counter-jet) is launched only
when its axis is tilted towards the companion star when the stars
are at periastron. When the accretion disk is nearly
edge-on as seen from the companion star, both the jet and the
counter-jet are launched. In practice, after running several tests we
determined that the accretion disk can be considered edge-on when
its axis is tilted with respect to the line joining the stars by
angles larger than 87.5\degr (i.e., approaching an orbital
configuration that is analogous to the Earth and Sun at solstice if
the polar axis of the Earth is thought as the axis of the disk).  A
scheme of the assumed configuration is shown in Figure \ref{fcroquis}.

\section{Initial setup of the numerical simulation}

The 3D numerical simulations were performed with the {\sc Yguaz\'u-a}
code \citep{raga00}. This code integrates the gasdynamical equations
with a second order accurate scheme (in time and space) using the
``flux-vector splitting'' method of \citet{vanleer82} on a binary
adaptive grid. Five levels of refinement were employed. Together with
the gas-dynamic equations, several rate equations for atomic/ionic
species were also integrated \citep[for details about the
  species used, reaction equations, and cooling rates,
  see][]{raga02}. 

The initial setup of the simulations is similar to that
  employed by \citet{riera14}. The main differences are: (1) in the
  present work we impose an asymmetrical jet ejection, which was not
  studied in \citet{riera14}; and (2) the jet is emitted at its
  maximum velocity when the system pass through periastron (see
  Eq.(\ref{eqvjet})).

At the center of a computational domain that spans (1.5, 1.5,
3)$\times 10^{17}$~cm (or $256\times 256\times 512$ cells at the
highest resolution of the grid) along the $x$--, $y$-- and $z$--axis,
respectively, a bipolar jet is imposed as a cylinder of radius $r_j$
and length $l_j$, both equal to $3.6\times 10^{15}$~cm (equivalent to
6 pixels in the finest grid). The jet axis rotates on the surface of a
precession cone with a half-opening angle $\alpha=15\degr$. The
precession period $\tau_p$ is four times the orbital period $\tau_o$,
which has a value of 15.4 yr. The jet velocity is given
by{\footnote{\citet{riera14} considered
    $v_j=v_{j0}(t)\times\left[1+\Delta v\ sin(\omega_o t)\right] ,$}}:

\begin{equation}
v_j=v_{j0}(t)\times\left[1+\Delta v \cos(\omega_o t)\right] ,
\label{eqvjet}
\end{equation}

\noindent where $\Delta v=0.5$, $\omega_o=2\pi / \tau_o$, and
$v_{j0}(t)$ is the mean jet velocity, which decreases linearly with
time at a rate of $-3\ \mathrm{km\ s^{-1}\ yr^{-1}}$ and has a
starting value of $400~\mathrm{km\ s^{-1}}$. The total number density
of the jet was fixed at $10^6~\mathrm{cm^{-3}}$. With these values,
the jet injects mass into the surrounding CSM at a rate
$\dot{M}_{\mathrm{j}}=5.5\times 10^{-5}\mathrm{M_{\sun}\ yr^{-1}}$ ($6.6
\times 10^{-3}\ \mathrm{M_{\sun}}$ are injected in 120~yr, the integration time of
our simulations).

At $t=0$~yr, the jet axis projected on the $xy$-plane forms an angle
of $-\pi/2$ with respect to the $x$--axis. Also, at this inital time,
the whole computational domain is filled with a slow and dense AGB
wind, with a density distribution given by \citet{mellema95}:

\begin{equation}
\rho(r,\theta)=\rho_w(r) \ f(\theta) ,
\label{rhow}
\end{equation}

\noindent where $\rho_w(r)$ is given by \citep{mellema95,riera14}:

\begin{equation}
\rho_w=\frac{1}{2}\bigg[(\rho_{\mathrm{sup}}+\rho_{\mathrm{AGB}})+(\rho_{\mathrm{sup}}-\rho_{\mathrm{AGB}})
  \cos\epsilon\bigg] \bigg(\frac{r_w}{r}\bigg)^2,
\label{rhoagb}
\end{equation}
where
\begin{equation}
\epsilon=\pi\times
\mathrm{min\bigg[1,\ max}\bigg[0,\ \frac{r-(r_\mathrm{w}+v_\mathrm{w}
      t_{\mathrm{sup}})}{v_\mathrm{w} t_{\mathrm{trans}}}\bigg]\bigg],
\label{fac_epsilon}
\end{equation}
$r$ is the distance from the primary star, $v_\mathrm{w}$ is
  the terminal wind velocity, and $r_\mathrm{w}$ is the stellar wind
  radius.  The times $t_\mathrm{sup}$ and $t_\mathrm{AGB}$ (see
  Eq. \ref{rhoagb}) indicate the duration of the superwind phase and
  the transition time between the AGB wind and the superwind
  phase, respectively; they were both chosen as 400~yr \citep{sanchez02,riera14}. The densities $\rho_\mathrm{AGB}$ and $\rho_\mathrm{sup}$ are calculated as:
\begin{equation}
\rho_\mathrm{AGB/sup}=\frac{\dot{M}_\mathrm{AGB/sup}}{4 \pi \ r_w^2 \ v_w}\, ,
\label{rhow2}
\end{equation}
where $\dot{M}_\mathrm{AGB}$ and $\dot{M}_\mathrm{sup}$ are the
  mass loss rates of the AGB and the super phase AGB winds,
  respectively.  We have chosen $\dot{M}_\mathrm{AGB}= 10^{-5}{\mathrm{ M_{\sun}\ yr^{-1}}}$, $\dot{M}_\mathrm{sup}= 10^{-4}{\mathrm{ M_{\sun}\ yr^{-1}}}$,
  $v_w=15\ {\mathrm{km\ s^{-1}}}$
  \citep{sanchez04,nakashima07,bujarrabal10,lee13}, $r_w=3.6\times
  10^{15}$~cm, and a constant temperature $T_w=100$~K.

Eq. (\ref{rhoagb})-(\ref{rhow2}) consider the mass-loss history of the
AGB star (i.e. we have considered the final AGB stage, in which the
star's $\dot{M}$ increases). The function $f(\theta)$ (see
Eq. \ref{rhow}) describes the angular dependence and can be written as

\begin{equation}
f(\theta) = 1-\delta \frac{1-\exp(-2 \beta \cos^2\theta)}{1-\exp(-2 \beta)},
\label{ftheta}
\end{equation}

\noindent with $\theta$ being the angle with respecto to the $z$-axis.
\noindent The parameters $\delta$ and $\beta$ where set to 0.7 and 5,
respectively. Thus, the equator-to-pole density ratio is given by
$1/(1-\delta)$. The parameter $\beta$ controls the shape of the density
distribution. The value chosen here yields the density distribution of a flat and dense
disk.

Furthermore, we have considered that the source of the jet describes
an elliptical orbit around the barycenter of the system with an
eccentricity of 0.5. Both foci of the orbit lie on the
$x$--axis. Since the orbit itself is not resolved by the simulation,
the effect of the orbital motion is to add a velocity component along
the orbit to the jet velocity.

By using the temperature and density distributions obtained from the
numerical simulations, we can compute the
[\ion{S}{2}]$\lambda\lambda6716$,$6730$ emission coefficients. The
intensities of these forbidden lines are calculated by solving
five-level atom problems, using the parameters of Mendoza (1983). The
[\ion{S}{2}]$\lambda\lambda6716$,$6730$ emission can be integrated
along lines of sight to produce synthetic maps.

\section{Results}

We have carried out hydrodynamical simulations in which the
asymmetrical jet ejection is either included (model M1) or not
included (model M2).  Model M2 is similar to that studied by
\cite{riera14} for the case $\tau_p/\tau_o=4$. We let the
hydrodynamical simulations of models M1 and M2 evolve until an
integration time of $120$~yr. At this time, the lobes of the simulated
pPN reach sizes similar to those observed in pPN
CRL~618 if a distance of $1$~kpc is assumed \citep{riera14}.

\subsection{Generating multipolar morphologies}

\begin{figure*}
\epsscale{1.5}
\plotone{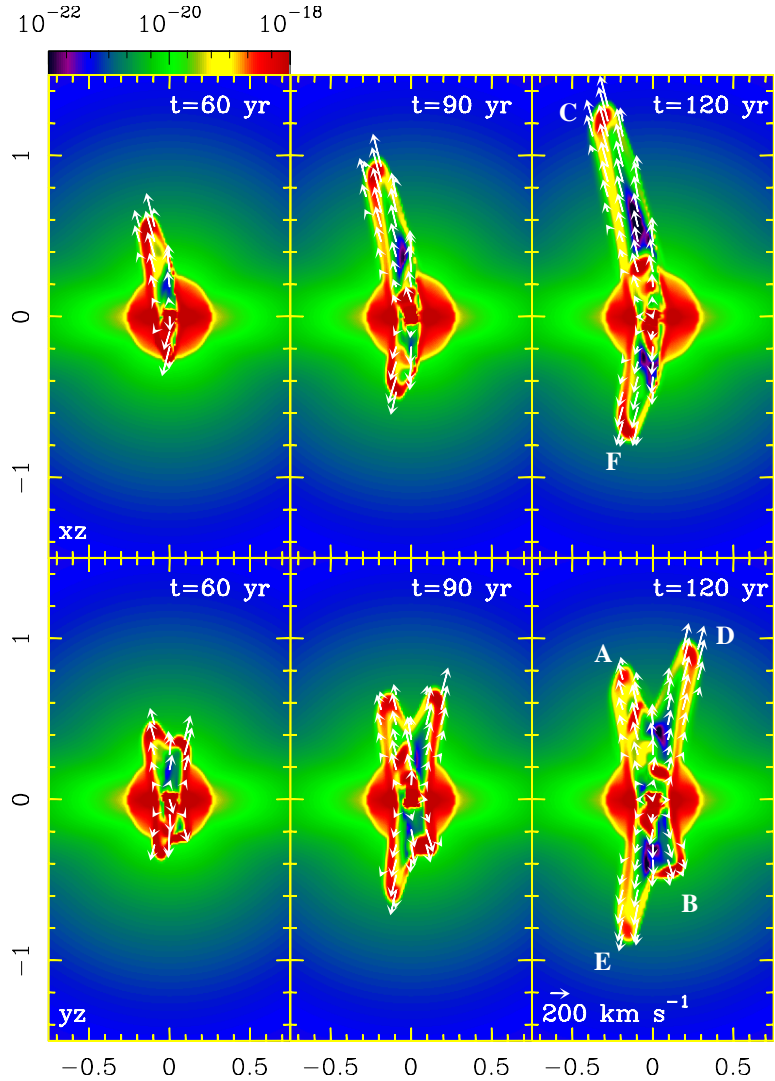}
\caption{Density stratification maps obtained at integration times of
  60, 90, and 120 yr. The upper (bottom) panels display the $xz$
  ($yz$) cut of the density stratification, passing through of the
  computational domain center, where the vertical axis corresponds to
  the $z$ direction. The white arrows show the velocity field of
  the flow. The horizontal arrow at the bottom left of the bottom
    right panel indicates a velocity of $200~\mathrm{km\ s^{-1}}$. The
    horizontal bar is the logarithmic color-scale of the density,
    which is given in units of $\mathrm{ g\ cm^{-3}}$. Both axes are
    displayed in units of $10^{17}$~cm. In order to facilitate
      the description of the density evolution, the lobes observed in
      both $yz$ and $xz$ projections (right panels) were labeled with
      capital letters from A to F (see text).
\label{fdens}}
\end{figure*}

Figure \ref{fdens} shows the temporal evolution of the density
  stratification of model M1 by means of density cuts on the $xz$- and
  $yz$-planes (passing through the center of the computational domain). 
  These sequences show the process through which a lobes distribution
  similar to that observed in the nebula CRL 618 can be reproduced. The flow velocity
  associated with the jet is displayed by white arrows; only velocities larger
than 50 $\mathrm{km\ s^{-1}}$ are shown.

In all $yz$ density maps (bottom panels of Figure \ref{fdens}), a
quadrupolar morphology is observed, exhibiting lobes with different
sizes.  Instead, the $xz$ density maps (upper panels of Figure
\ref{fdens}) only show a bipolar lobe morphology. The formation of these
structures can be understood by following the ``launch sequence of the jet'',
which is as follows: (1) at
$t=0$~yr the top-left and the bottom-right lobes (hereafter lobes A
and B, respectively) observed in the $yz$ density maps are launched;
(2) at $t=\tau_o$ (as was previously mentioned, the jet velocity
variability period is equal to the orbital period, $\tau_o$), the
top-left lobe (hereafter lobe C) observed in the $xz$ density map is
emitted; (3) at $t=2\tau_o$, both top-right and bottom-left lobes
observed (hereafter lobes D and E) in the $xz$ density map are
launched; (4) at $t=3\tau_o$ the bottom-left lobe (hereafter lobe F)
observed in the $yz$ density map is ejected. This sequence is then
repeated for subsequent times.

An interesting point to note is that the lobes A and B observed in the
$yz$ density maps (see Figure \ref{fdens}) have different sizes
despite being launched at the same time. This difference is caused by
the orbital motion of the jet source and the density distribution
around it. As the jet source passes trough periastron, the orbital
speed reaches its maximum value of $\sim 30
\mathrm{km\ s^{-1}}$. Although this value is small compared to the
initial jet launching velocity (of $600~\mathrm{km\ s^{-1}}$), its
vector addition is enough to cause a measurable difference in the
directions of motion of the jets of $\sim5\degr$ (see the white arrows
in Fig.\ref{fdens}).   Since the AGB wind material has the
density distribution of a dense disk parallel to the $xy$ plane, the
material of lobe B propagates in a denser medium than that of lobe A,
thus decelerating quicker and causing lobe B to be shorter than lobe
A. Similarly, the C lobe (shown in the $xz$ density maps) is the
larger one because it propagates into the cavity created by the
previous lobes, which has a lower density.  

\subsection{Synthetic maps: comparison with observations}

\begin{figure}
\epsscale{1.}  \plotone{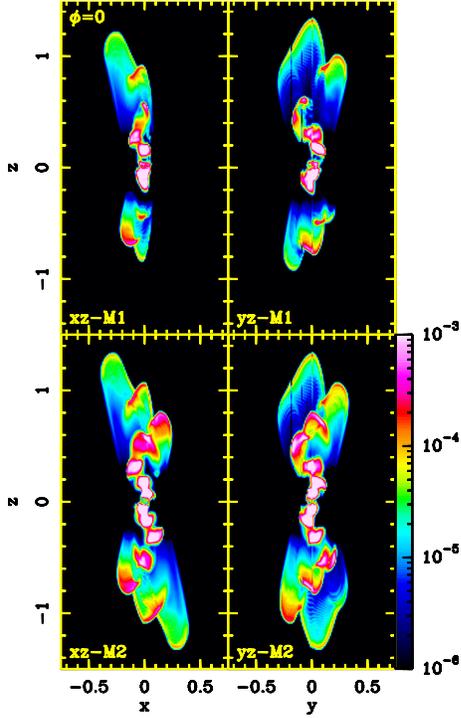}
\caption{Synthetic [\ion{S}{2}] emission maps obtained from the
  numerical simulations. The upper panels display the $xz$ (left) and
  $yz$ (right) projections for the model M1.  The $z$--axis of the
  computational domain lies on the plane of the sky ($\phi=0$). The
  bottom panels display the same but for model M2. Both axes are given
  in units of $10^{17}$~cm. The vertical color bar indicates the
  [\ion{S}{2}] emission in units of $\mathrm{erg\ s^{-1}
    \ cm^{-2}\ sr^{-1}}$.
\label{f1}}
\end{figure}

With this launching sequence of the jet, we expect to obtain
morphologies similar to that of CRL~618 if we ``observe'' the system
along the $y$ direction, i.e., when the line of sight is aligned with
the $y$ axis. In order to test this, synthetic [\ion{S}{2}] emission
maps (shown in Figure \ref{f1}) were obtained from the numerical
results of models M1 and M2 considering different lines of
sight. These maps were generated considering that the precession axis
lies on the plane of the sky ($\phi=0\degr$). The top panels of Figure
\ref{f1} show the $xz$ and $yz$ projections (left and right panels,
respectively) for model M1, while the bottom panels show the
corresponding map projections for model M2. The main differences
between these models are evident by comparing the morphology of the
two maps obtained for the $xz$ projection. On the one hand, the
morphology obtained for model M2 shows six lobes, while four lobes are
observed for model M1. On the other hand a clear point-symmetric lobe
distribution is observed in the map of the model M2, while the map for
model M1 displays an almost mirror-symmetric ``nebula'', which
resembles the observed morphology of the pPN CRL~618. The size and
orientation of the upper lobes of the synthetic nebula are similar to
the observed Eastern lobes of the pPN CRL~618. A similar agreement is
observed for the bottom lobes. In addition, the $xz$ projection map of
model M1 (see Figure \ref{f1}) reveals a remarkable difference in the
morphology and the size of the lobes corresponding to the jet and the
counter-jet.

\begin{figure}
\epsscale{1.}
\plotone{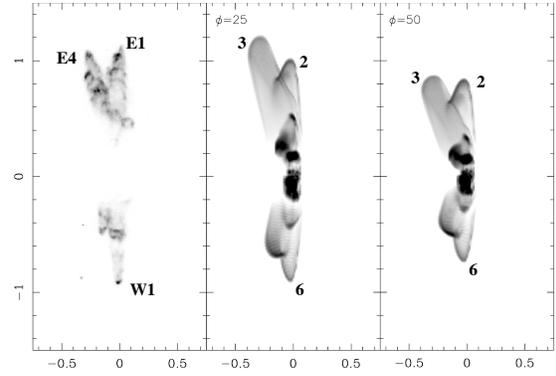}
\caption{Direct comparison between the observed (left panel) and
  synthetic [\ion{S}{2}] images (middle and right panels). The
  synthetic [\ion{S}{2}] maps ($xz$ projection) were obtained considering that the
  precession axis is tilted by either 25\degr or 50\degr (middle and
  right panels, respectively) with respect of the plane of the
  sky. The horizontal and vertical axes are given in units of
  $10^{17}$~cm. The three more evolved fingers have been labeled in order to
  facilitate the comparison (as described in the text).
\label{fcomp}}
\end{figure}

A direct comparison between observations and synthetic emission
  maps is shown in Figure \ref{fcomp}. The observed [\ion{S}{2}] image
  of the pPN CRL 618 is displayed on the left, while the middle and
  the right panels show the $xz$ projection of the synthetic emission
  map considering angles between the precession cone axis and the
  plane of the sky of 25\degr (middle) and 50\degr (right). The
  orientation of the observed nebula with respect to the plane of the
  sky is not precisely determined, with estimates ranging from 20\degr
  to 40\degr (S\'anchez-Contreras et al. 2002).  In order to facilitate
  the comparison we have labeled three fingers in the observed image
  as E1, E4 and W1, following the labeling given by
  \citet{balick13}. Also, on the synthetic maps, several ``bow shock''
  features were labeled with numbers from 1 to 6. Synthetic maps
  reveal a four-fingered structure. In order to identify these
  fingers, we employ the number of the bow shock features located at
  the finger tip.  Finger ``2'' is the overlap, along the line of
  sight, of the lobes labeled as ``A'' and ``E'', which were mentioned
  in the previous subsection, while finger ``6'' is the overlap of
  lobes ``B'' and ``E''. Fingers ``3'' and ``5'' correspond to lobes
  ``C'' and ``F'', respectively. It must be noted that while the
  overall size of the ``synthetic nebula'' for the 25\degr case
  (middle panel of Figure \ref{fcomp}) is similar to the observed one
  (if a distance of 1~kpc is considered), the sizes of fingers ``2''
  and ``3'' are dissimilar, in contrast to the observations. The size
  of fingers ``2'' and ``3'' becomes similar if the angle is 50\degr
  (right panel). However, in this case, the ``synthetic nebula'' is
  somewhat smaller than the observed one for the same assumed
  distance.

\subsection{Proper motions and PV diagram}

\begin{figure}
\epsscale{1.}  \plotone{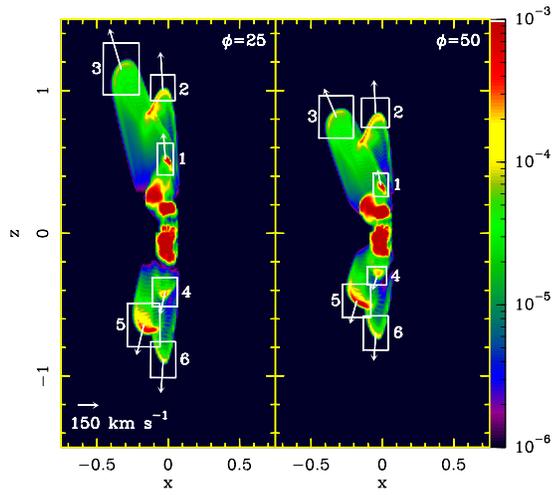}
\caption{Proper motions of the individual features (labeled from 1 to
  6) obtained for the [\ion{S}{2}] synthetic maps (model M1) at
  $\phi=25$\degr (left panel) and $\phi=50$\degr (right panel). The
  white boxes show the regions that were used to perform the proper
  motion study. The arrows indicate the velocities of the features in
  the plane of the sky (see Table \ref{tab:models}). The horizontal white arrow at the bottom left
  of the left panel indicates a proper motion of
  $150~\mathrm{km\ s^{-1}}$. The logarithmic color scale gives the
  [\ion{S}{2}] flux in units of $\mathrm{erg\ s^{-1}
    \ cm^{-2}\ sr^{-1}}$.
\label{fpm}}
\end{figure}

Employing the synthetic maps, we carried out a proper motion
  study. As in \citet{riera14}, the simulations were restarted from
  the output obtained at an integration time of 120~yr and they were
  left to evolve 10~yr longer. The two ouputs are then compared to one
  another. The results obtained from this proper motion study are
  shown in Figure \ref{fpm} and summarized in Table
  \ref{tab:models}. The main resul is that the kinematic ages of
  fingers ``2'', ``3'', and ``6'' found by this analysis turn out to
  be very similar to each other ($\sim 100$~yr) for both orientations
  with respect to the plane of the sky that we have considered. This is in good
  agreement with the kinematic ages obtained by \citet{balick13} for
  fingertips E1, E4, and W1, determined by using both F606W and F656N
  images of the pPN CRL~618.

This result could look like striking because actually finger
  ``3'' was ejected 15 yr after the launching of fingers ``2'' and
  ``6'' (see subsection 4.1). This is, we would expect that the finger
  3 turns out to be 15 yr younger than fingers ``2'' and
  ``6''. However it is necessary to take into account that the
  material associated to fingers ``2'' and ``6'' do not follow a
  ballistic trajectory because at the beginning they move into a dense
  CSM. Instead, the gas of finger ``3'', which was launched later,
  propagates into the lower density cavity which was previously
  excavated by the material belonging to fingers ``2'' and ``6'' (or
  lobes ``A'' and ``B'', see subsection 4.1), thus evolving faster.

\begin{figure}
\epsscale{1.}
\plotone{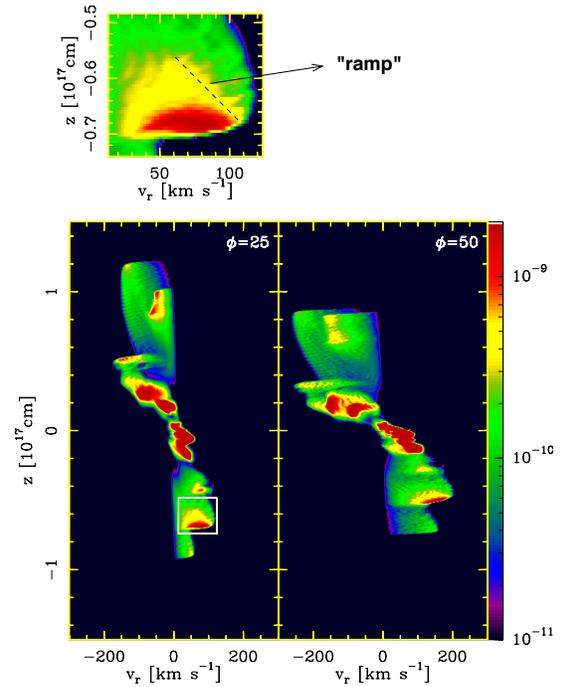}
\caption{PV diagrams obtained for model M1 ($xz$ projection)
  considering angles between the precession axis of the jet and the
  plane of the sky of 25\degr (left panel) and 50\degr (right
  panel). The vertical logarithmic color scale gives de flux in units
  $\mathrm{erg\ s^{-1} \ cm^{-2}\ sr^{-1}}$.  The upper panel is an
  enlargement of the rectangular white region in the left panel.
\label{f3}}
\end{figure}

Figure \ref{f3} displays PV diagrams (bottom panels) for model M1
($xz$ projection) considering both orientations with respect to the
plane of the sky. As in the Figure 3 of \citet{riera11}, the PV
diagrams display several ramps in radial velocity. This is clearly
shown in the enlargement (upper panel) of the white rectangular region
in the left panel. Furthermore, the jet ejection mechanism employed in
this work produces clear differences in both the emission and the
radial velocities of the blue- and the red-shifted components. The
radial velocities differ by factors which lie in the range 1.3--2,
which is consistent with the values reported by \citet{hirth94}.

\section{Discussion and conclusions}

As was mentioned above, it is not uncommon to find asymmetries in the
jet velocities of HH objects, with factors between the blue and red
components of about 1.5--2.5 \citep{hirth94}. The nature of these
asymmetries can be intrinsic, which is related to the ejection
mechanism itself, or extrinsic, where the properties of the CSM
determine the shapes and sizes of lobes and their
expansion velocities \citep{matsakos12}. It can be
expected that the asymmetries observed in jets of pPNe share
the same nature as those seen in the jets of HH objects.

Recently, \citet{steffen13} reproduced multipolar morphologies by
means of the interaction of a fast isotropic wind with a very
inhomogeneous CSM, an idea that can be classified under the
  extrinsic origin possibility.

In contrast to this, the ``twin-jets'' mechanism proposed by
  \citet{soker13} belongs to the intrinsic class. This mechanism is
  based on the existence of a binary system at the center of the
  object, and it consists in that two narrow jets are
  emitted at the same time from a narrow region when the jet source passes
  through periastron. With this scenario, they explain the morphology observed
  in the pPN CRL618 and in the YSO NGC 1333 IRAS 4A2.

The asymmetric jet ejection mechanism that we propose in this work
also belongs to the intrinsic origin hypothesis.  We study the effects
of an imposed alternation (between the two outflow lobes) of the
ejection on the nebular morphology.  Depending on the orientation of
the system with respect to the observer, large scale morphologies with
a mirror symmetry are obtained, such as that observed in the pPN
CRL~618. The main observational kinematic features such as PV diagrams
\citep{riera11} and proper motions \citep{riera14} are
recovered. Also, the PV diagram corresponding to model M1 shows
differences in the radial velocities between the blue and the red
components by factors of 1.3--2, which is in agreement with the result
of \citet{hirth94}, for the case of HH objects. Furthermore, the
  kinematic ages of the fingers obtained from our numerical study are
  in a very good agreement with those reported by \citet{balick13}.

As we have pointed out in the past \citep{masciadri02,raga09}, an
orbital motion of the jet source can result in mirror symmetries of
the outflow lobes.  However, the relatively low orbital velocities
possible for the sources of observed mirror-symmetric pPNe cannot be
responsible for the large angular deviations of their outflow lobes
\citep{vel13,riera14}.

In the simulations presented above, we show that a jet ejection episodes
that alternate between the two sides of the accretion disk (synchronized with the
precession of the disk) can indeed produce the observed
mirror-symmetric multi-lobe structures. This presents a very interesting alternative to the ``twin-jet'' model for explaining the mirror-symmetric lobes with large angular deviations observed in some PPNe.

\acknowledgments The authors acknowledge an anonymous referee for her/his very fruitful and useful comments, which helped us improve the previous version of this manuscript.  PFV, JCTR and ACR thank financial support from CONACyT grants 167611 and 167625, and DGAPA-PAPIIT (UNAM) IG100214. A.Ri. is partially supported by Spanish MCI grants AYA2008-06189-C03, AYA2011-30228-C03, and FEDER funds. We also thank Enrique Palacios for maintaining the computer cluster on which the simulations were carried out.


\clearpage

\clearpage

\begin{deluxetable}{cccrrrr}
 \tablecolumns{7} \tablewidth{0pc} \tablecaption{Proper motion
   velocities corresponding to Figure \ref{fpm} 
\label{tab:models}}
 \tablehead{ \colhead{Feature} & \colhead{$\phi$} & \colhead{Distance} & \colhead{$v_x$} &
   \colhead{$v_z$} & \colhead{$v_t$} & kinematic age\\ \colhead{}& \colhead{[\degr]} &
   \colhead{[$10^{17}$~cm]}& \colhead{[$\mathrm{km\ s^{-1}}$]}
   &\colhead{[$\mathrm{km\ s^{-1}}$]} &
   \colhead{[$\mathrm{km\ s^{-1}}$]} & \colhead{[yr]} } 
\startdata 1 &  25.0 & 0.50 & -15.0 & 210.0 & 211.0 & 79.0 \\ 
           1 &  50.0 & 0.34 & -24.0 & 122.0 & 124.0 & 87.0 \\ 
           2 &  25.0 & 1.00 & -18.0 & 303.0 & 304.0 & 108.0 \\ 
           2 &  50.0 & 0.85 & -8.9 & 262.0 & 262.0 & 102.0 \\ 
           3 &  25.0 & 1.20 & -93.0 & 322.0 & 335.0 & 114.0 \\ 
           3 &  50.0 & 0.88 & -95.0 & 228.0 & 247.0 & 113.0 \\ 
           4 &  25.0 & 0.40 & -39.0 & -178.0 & 182.0 & 72.0 \\ 
           4 &  50.0 & 0.30 & -34.0 & -113.0 & 118.0 & 81.0 \\ 
           5 &  25.0 & 0.70 & -55.8 & -246.0 & 252.0 & 84.0 \\ 
           5 &  50.0 & 0.51 & -50.0 & -181.0 & 188.0 & 86.0 \\ 
           6 &  25.0 & 0.90 & -18.0 & -266.0 & 267.0 & 106.0 \\ 
           6 &  50.0 & 0.70 & -14.0 & -221.0 & 222.0 & 100.0 \\ 
\enddata
\end{deluxetable}

\clearpage


\end{document}